\documentclass[12pt,english]{emulateapj}
\usepackage[latin1]{inputenc}
\usepackage{graphicx}
\usepackage{amssymb}
\usepackage{esint}

\makeatletter

\providecommand{\tabularnewline}{\\}

\@ifundefined{definecolor}
 {\usepackage{color}}{}
\usepackage{array}

\makeatletter

\usepackage{times}

\definecolor{DarkBlue}{rgb}{0.1,0,0.6}

\newcommand{\pc}{\mathrm{pc}}\newcommand{\msun}{M_\odot}\newcommand{\kms}{\mathrm{km\,s^{-1}}}

\shorttitle{Dynamical ejection of runaway stars}\shortauthors{Perets \& Subr}

\makeatother
\usepackage{babel}

\makeatother

\usepackage{babel}

\makeatother

\usepackage{babel}

\makeatother

\usepackage{babel}

\begin{document}

\title{The properties of dynamically ejected runaway and hyper-runaway stars}

\author{Hagai B. Perets$^{1}$ and Ladislav \v{S}ubr$^{2}$}

\affil{$^{1}$Harvard-Smithsonian Center for Astrophysics, 60 Garden St.,
Cambridge, MA, USA 02138\\
 $^{2}$Astronomical institute, Faculty of Mathematics and Physics,
Charles University, V Hole\v{s}ovi\v{c}k\'ach 2, 18000 Praha, Czech
Republic}
\begin{abstract}
Runaway stars are stars observed to have large peculiar velocities.
Two mechanisms are thought to contribute to the ejection of runaway
stars, both involve binarity (or higher multiplicity). In the binary
supernova scenario a runaway star receives its velocity when its binary
massive companion explodes as a supernova (SN). In the alternative
dynamical ejection scenario, runaway stars are formed through gravitational
interactions between stars and binaries in dense, compact clusters
or cluster cores. Here we study the ejection scenario. We make use
of extensive N-body simulations of massive clusters, as well as analytic
arguments, in order to to characterize the expected ejection velocity
distribution of runaways stars. We find the ejection velocity distribution
of the fastest runaways ($v\gtrsim80\,\kms$) depends on the binary
distribution in the cluster, consistent with our analytic toy model,
whereas the distribution of lower velocity runaways appears independent
of the binaries properties. For a realistic log constant distribution
of binary separations, we find the velocity distribution to follow
a simple power law; $\Gamma(v)\propto v^{-8/3}$ for the high velocity runaways and  $v^{-3/2}$ for the low velocity ones. We calculate the
total expected ejection rates of runaway stars from our simulated
massive clusters and explore their mass function and their binarity.
The mass function of runaway stars is biased towards high masses,
and depends strongly on their velocity. The binarity of runaways is
a decreasing function of their ejection velocity, with no binaries
expected to be ejected with $v>150\,\kms$. We also find that hyper-runaways
with velocities of hundreds of $\kms$ can be dynamically ejected from
stellar clusters, but only at very low rates, which cannot account
for a significant fraction of the observed population of hyper-velocity
stars in the Galactic halo. 
\end{abstract}

\section{Introduction}

Runaway stars are those stars observed to have large peculiar velocities
($40\le v_\mathrm{pec}\le200\,\kms$ \citep{bla61,gie87,hoo+01} and even
higher \citep{mar06}; the specific definitions vary). A considerable
fraction of the early O and B stars population are known to be runaway
stars ($\sim30-40\%$ of the O stars and $5-10\%$ of the B stars;
\citealt{sto91} and refs. therein). The velocity dispersion of the
population of runaway stars is much larger than that of the `normal'
early-type stars \citep{sto91}. Besides their relatively high velocities,
runaway stars are also distinguished from the normal early-type stars
by their much lower ($<10\%$) multiplicity compared with the binary
fraction of normal early-type OB stars ($>50\%$ and up to $100\%$;
\citealt{gar+80,mas+98,kob+07,kou+07}).

Two mechanisms are thought to contribute to the acceleration of regular
runaway stars, both involve binarity (or higher multiplicity). In
the binary supernova scenario (BSS; \citealt{bla61}) a runaway star
receives its velocity when the primary component of a massive binary
system explodes as a supernova (SN). When the SN shell passes the
secondary the gravitational attraction of the primary reduces considerably,
and the secondary starts to move through space with a velocity comparable
to its original orbital velocity. In the dynamical ejection scenario
(DES; \citealt{pov+67}) runaway stars are formed through gravitational
interactions between stars in dense, compact clusters. Simulations
show that such encounters may produce runaways with velocities up
to $200\,\kms$, and even higher in more rare cases (\citealt{mik83,leo+90,leo+91,gua+04}).
These scenarios suggest that many of the early OB stars formed in
young clusters could be ejected from their birth place and leave the
cluster at high velocity. Interestingly, observations show that even
very massive O-stars could be accelerated to become runaways \citep{com+07}.

In recent years, stars with extremely high peculiar velocities (hyper-velocity
stars; HVSs) of a few$\times10^{2}\kms$ have been observed in the
Galactic halo \citep{bro+05,bro+06a,bro+06b,bro+07,ede+06}. HVSs
are thought to be ejected from the Galactic center following a dynamical
interaction with the massive black hole (MBH) known to exist in the
center \citep{hil88,han+03,yuq+03,gua+05,bau+06,bro+06,lev06,per+07,per09b}. Nevertheless, it was suggested
that the BSS and/or the DES could eject, under some conditions, stars
with extreme velocities (termed hyper-runaways), comparable with those
of observed HVSs, possibly explaining the origin of some of these
HVSs \citep{gva+09}.

Several studies have explored the DES (see \citealp{hoo+00}) using
N-body simulations. Most of these have focused on single encounters
between a single/binary star and another binary star, and found the
velocities of the ejected stars in such encounters \citep[e.g.][]{gva+09}.
\cite{leo+90,leo+88} explored the dynamical properties of runaway stars
and their distribution in a stellar cluster environment, and not in single
encounters. Their studies, however, were limited to a relatively
small number of simulation of very small clusters ($\sim30$ stars;
in addition to some hybrid N-body - Monte-Carlo simulations of larger
clusters of a few hundreds of stars), and hence to very small sample
of ejected runaway stars. Recently, \citet{tan+09}, \citet{fuj+11} and \citet{ban+12} studied the ejection of stars from large clusters. These simulations provide important progress on these issues, and complement the current study, but they include only a limited number of simulations, which do not provide enough statistics of the high velocity tail distribution of runaways and hyper-runaways, and can not be used for the analysis done here. 
Note that \citet{fuj+11} study considered only dynamically
formed binaries, and did not include primordial binaries in their
simulations. \citet{ban+12} studied extremely massive and more clusters ($N=10^5$ stars). These approaches explore different regimes and/or processes than explored by us, and complement our current study.  

Here we develop an analytic understanding of velocity distribution
of the fastest runways in the DES. We compare it with a large sample
of simulated runaway stars produced in extensive N-body simulations
of hundreds of large stellar clusters. These provide us, for the first
time, large enough database of runaway stars with high velocities,
which could be analyzed statistically. Using our N-body simulations
we characterize the ejection rates, velocity distribution, and binarity
of dynamically ejected runaway stars. We find cases of stars dynamically
ejected at extreme ejection velocities of a few hundreds of $\kms$,
however these are relatively rare cases and are not likely to explain
the vast majority of hyper-velocity stars observed and inferred to
exist in the Galactic halo.

\section{Analytical estimates}

We begin by exploring analytically the dynamical ejection scenario
for runaway and hyper-runaway stars. We consider an interaction of
a binary of mass $M_{\mathrm{B}}=M_{1}+M_{2}$ and a single
star $M_{\star}$. Large accelerations of the single star are
assumed to occur when it passes one of the binary components within
the semi-major axis, $a$, of the binary. Assuming that the scattering
is dominated by the gravitational focusing, the cross section of the
interaction is \begin{equation}
\sigma(a)\approx\frac{2\pi GM_{{\rm B}}a}{v_{{\rm c}}^{2}}\;,\label{eq:cross-section}\end{equation}
 where $v_{\mathrm{c}}$ is the characteristic stellar velocity in
the cluster.

The energy exchange between a hard binary and star is comparable to
the binary orbital energy. Hence, the energy transfer to the star
is of the order \begin{equation}
\Delta E_{\star}\approx\frac{G M_\mathrm{B}M_{\star}}{a}\;.\end{equation}
 and the star acquires a large velocity kick, $v_\mathrm{kick}$ of the order
of the orbital velocity of the binary. We are interested in the cases
where $\Delta E_{\star}$ is much larger that the star's energy before
the interaction. The velocity of the ejected star $v_\mathrm{kick}\simeq\sqrt{\Delta E}$
is therefore estimated as \begin{equation}
v_{\mathrm{esc}}\approx\left(\frac{2G M_{\mathrm{B}}}{a}\right)^{-\alpha}\;,\label{eq:vesc}\end{equation}
 with $\alpha=1/2.$ Simulations of binary-single star encounters
suggests a somewhat steeper slope of $\alpha\sim3/5$ \citep[see figure 4 in][]{gva+09}.

The differential cross section per unit volume of the binary--single
star interaction with semi-major axis in the interval $\langle a,a+\delta a\rangle$
is \begin{equation}
\frac{dR(a)}{da}\approx n(a)N_{\mathrm{B}}n_{\star}\sigma(a)v_{c}\label{eq:dRa}\end{equation}
 where $n_{\mathrm{B}}$ is the number of binaries per unit volume,
$n_{\star}$ is number density of stars and $n(a)$ is the differential
distribution of semi-major axis of the binaries, which is normalized
to unity: \begin{equation}
\int_{a_{{\rm min}}}^{a_{{\rm {\rm max}}}}da\, n(a)=1\;.\end{equation}

Let us consider a log constant distribution of the binaries semi-major
axis (so called $\ddot{{\rm O}}$pik's law, with $n(a)\propto a^{-\beta}$;
$\beta=1$), representing an empirical distribution of massive binaries
\citep[e.g.][]{kob+07} and, for testing purposes, an uniform distribution
$n(a)={\rm \mathrm{const}}$; $\beta=0$. Due to the gravitational
focusing, the cross section of encounters with massive binaries is
much higher than with low mass stars. In addition, the ejection velocities
from encounters with massive stars are higher. Observationally, massive
stars also have a much higher binarity fraction, as well as separation
distribution which is biases toward closer binaries (lower mass stars
seem to have a log normal distribution of periods; \citealp{duq+91}).
For all these reasons, the majority of runaway stars are ejected through
interactions of massive stars with massive binaries. For simplicity,
we neglect dependence of $dR(a)$ on the specific mass function of
binaries in Eq.~(\ref{eq:dRa}), and assume some fiducial effective
typical mass for the stars involved in the dynamical encounter. Interestingly,
our numerical simulation results suggest that this may be justified
even for a range of stellar masses, however, we leave the more detailed
analytic study of this dependence for further investigation.

Eq. (\ref{eq:dRa}) can be directly transformed to the production
rate of high velocity stars per unit volume and velocity interval
$dv_{\mathrm{esc}}$: \begin{equation}
\frac{dR(v_{\mathrm{esc}})}{dv_{\mathrm{esc}}}\approx4GM_{\mathrm{B}}v_\mathrm{esc}^{-(\alpha+1)/\alpha}n(a)n_{\mathrm{B}}n_{\star}v_{\mathrm{c}}\,\sigma(a)\propto v_\mathrm{esc}^{-\left(\frac{2+\alpha-\beta}{\alpha}\right)}\;.\label{eq:rate}\end{equation}
 Making use of Eqs. (\ref{eq:cross-section}) and (\ref{eq:vesc})
we can find the expected velocity distribution of the ejected stars,
for some appropriate choice of the binaries separation distribution.
For an Opik's distribution of the semi-major axis $n(a)\propto a^{-1}$
we expect the velocities of ejected stars to be distributed as $R(v_{{\rm esc}})\propto v_{{\rm esc}}^{-8/3}$
($\alpha=3/5$; $\beta=1$), while for $n(a)=\mathrm{{\rm const}.}$
we get $R(v_{{\rm esc}})\propto v_{{\rm esc}}^{-13/3}$ ($\alpha=3/5$;
$\beta=0$). This approach can be naturally extended to any desired
distribution of binary separations.


\section{N-body simulation models}
{
\begin{table*}
\begin{centering}
\begin{tabular}{c|cccccccccc}
\scriptsize 
model  & \scriptsize $N_{\mathrm{run}}\,^1$  & \scriptsize $r_{{\rm h}}\,(pc)\,^2$  & \scriptsize $T_{\mathrm{max}}(Myr)\,^3$  & \scriptsize $S\,^4$  & \scriptsize $\!\! M_{\mathrm{c}}\,(M_{\odot})\,^5$  & \scriptsize IMF $^6$  & \scriptsize $n(a)\,^7$  & \scriptsize $N_{\mathrm{bin}}\,^8$  & \scriptsize $M_{\mathrm{p}}\,(M_{\odot})\,^9$  & \scriptsize $M_{\mathrm{s}}\,(M_{\odot})\,^{10}$ \tabularnewline
\hline 
\scriptsize 009  & \scriptsize 200  & \scriptsize $0.2$  & \scriptsize $2.7$  & \scriptsize 0.25  & \scriptsize $5000$  & \scriptsize Salpeter  & \scriptsize $a^{-1}$  & \scriptsize $\approx73$  & \scriptsize $>4$  & \scriptsize $>1$ \tabularnewline
\scriptsize 014  & \scriptsize 500  & \scriptsize $0.2$  & \scriptsize $4.4$  & \scriptsize 0.25  & \scriptsize $5000$  & \scriptsize Salpeter  & \scriptsize $\mathrm{const.}$  & \scriptsize $\approx73$  & \scriptsize $>4$  & \scriptsize $>1$ \tabularnewline
\hline 
\scriptsize 016  & \scriptsize 200  & \scriptsize $0.4$  & \scriptsize $35.5$  & \scriptsize 0  & \scriptsize $10000$  & \scriptsize $200\times10\, M_{\odot}$  & \scriptsize $a^{-1}$  & \scriptsize 100  & \scriptsize $10$  & \scriptsize $10$\tabularnewline
  &   &  &  &  &  & \scriptsize $8000\times1\, M_{\odot}$  &  &  &  & \tabularnewline
\scriptsize 017  & \scriptsize 200  & \scriptsize $0.4$  & \scriptsize $35.5$  & \scriptsize 0  & \scriptsize $10000$  & \scriptsize $200\times10\, M_{\odot}$  & \scriptsize $\mathrm{{\rm const.}}$  & \scriptsize 100  & \scriptsize $10$  & \scriptsize $10$ \tabularnewline
  &   &  &  &  &  & \scriptsize $8000\times1\, M_{\odot}$  &  &  &  & \tabularnewline
\end{tabular}
\par\end{centering}
\caption{\label{tab:models}}
{\small  Parameters of the models: (1) Number of simulation runs. (2) initial half-mass radius $r_{\mathrm{h}}$. (3)  Simulation time. (4) Index of the initial mass segregation $S$ (5) Total mass of the cluster $M_{\mathrm{c}}$ (6) Mass function. (7) Distribution of the binary semi-major axis $n(a)$. (8) Initial percentage of binaries $f_{\mathrm{bin}}$. (9) Mass of the primary $M_{\mathrm{p}}$ and (10) secondary $M_{\mathrm{s}}$ star. Salpeter initial mass function has power-law profile $\propto M_{\star}^{-2.35}$
within the interval $\langle0.2\, M_{\odot},80\, M_{\odot}\rangle$.
In the models 009 and 014, physical collisions of stars are allowed
(i.e. the mass function is not constant). Initially, the binaries
have semi-major axes in the range $\langle0.05\,{\rm {\rm AU}},50\,{\rm {\rm {\rm AU}}}\rangle$
and zero eccentricities.}
\end{table*}
}

We have studied several different numerical models of star clusters
in order to characterize the properties of runaway stars. We also
make use of simplified cluster models to verify our analytic calculations
of the the velocity distribution of ejected runways. Table~\ref{tab:models}
summarizes basic physical parameters of the models which were integrated
by means of NBODY6 code \citep{aar99}. In all cases, distribution
of the binary semi-major axis was truncated outside the interval $\langle0.05\mathrm{\,{\rm AU}},50\,\mathrm{{\rm AU}}\rangle$;
initial eccentricities were set to zero. In the first models (denoted
016 and 017 below) we studied
a highly simplified stellar systems in which only two masses were used
rather than a continuous mass function. All binaries in these models
are made of identical $10\, M_{\odot}$ components and the single
stars are identical $1\, M_{\odot}$ stars. In addition, all stars
are treated as point-masses in these models, i.e. the models are scale-free.
In Table~\ref{tab:models} we present scaling with initial half-mass radius
$r_\mathrm{h} = 0.4\,\pc$ which, together with considered stellar masses,
implies scaling of time. The models were integrated to $35.5\,\mathrm{Myr}$.
Later we also compared
these to simulations results of similar clusters with no primordial
binaries, to verify that single-single encounters can only lead to
ejections with $v_{{\rm esc}}\lesssim20\,\kms$. During the cluster
evolution, dynamically formed binaries could eject stars at higher
velocities, but the overall fraction of runaways was only a small
fraction than in the case of clusters including primordial binary
population. 

Though in this study we focus on the simple cluster simulations, we
also run simulations of somewhat more realistic clusters, in which
the full ranges of Salpeter initial mass function (IMF) for the stars
is considered. In the latter models (009 and 014), all massive
($M\ge4\, M_{\odot}$)
stars reside in primordial binaries, and we set a lower limit for
the primary and the secondary star to $4\,M_{\odot}$ and $1\, M_{\odot}$,
respectively. We use pairing algorithm which prefers similar masses
of the two components which is in accord with observations.
More specifically, the algorithm first sorts the stars from the
most massive to the lighest one. The most massive star from the set is taken as
the primary. The secondary star index, $id$, in the ordered set is generated as a
random number with the probability density $\propto id^{-\beta}$ and $\max(id)$
corresponding to a certain mass limit, $M_\mathrm{s,min}$. The two stars are
removed from the set and the whole procedure is repeated until stars with
$M_\star \geq M_\mathrm{p,min}$ remain. We used as the minimal mass
of the primary $M_\mathrm{p,min} = 4\msun$, as the minimal mass of the secondary
$M_\mathrm{s,min} = 1\msun$ and the pairing algorithm index $\beta=40$.
Furthermore,
binary stars in real clusters may physically collide. Hence, we enable
the possibility of stellar collisions for these models in order to
increase their realism; stars are merged if they pass to each other
at a distance smaller than the sum of their radii. Finite stellar radii
(we adopted simple relation $R_\star=R_\odot (M_\star / M_\odot)$ and 
$R_\star=R_\odot (M_\star / M_\odot)^{4/5}$ for $M_\star \leq M_\odot$ and
$M_\star > M_\odot$ respectively) establish scales within the clusters. Both
models 009 and 014 have total mass of $5000\,M_\odot$ ($\sim7400$ stars for a Salpeter IMF in the given mass range) and the initial
half-mass radius $r_\mathrm{h} = 0.2\, \pc$. They have been integrated to
$T=2.7\,\mathrm{Myr}$ and $4.4\,\mathrm{Myr}$ respectively. Nevertheless, we use results only from the first 2.7 Myrs of evolution, such that the ejections from the different clusters could be directly compared. 

Note that we do not include stellar evolution is our models, as even the most massive stars in our simulation have a longer main sequence lifetime, and therefore stellar evolution does not play a role. Models 016 and 017 are followed for longer timescale (35.5 Myrs), but these are idealized two-mass models, serve to explore the overall dynamical processes and not the overall realistic evolution.   
Evolution of stellar clusters over longer timescales would be affected by stellar evolution. In particular supernovae explosions could produced runaways through the BSS scenario. Here we focus on the DES case and do not explore the longer time evolution in which stellar evolution can play an important role. 

Initial state of the models 009 and 014 correspond to mass segregated state according to \cite{sub+08}. Briefly, this setup is based on an empirical finding that
mean specific binding energy of stars in numerical models of star clusters
tend to a power-law relation to stellar masses.  Profiles of initially mass
segregated models cannot be expressed analytically but, in general, their
density increases towards the center. All models under consideration were
assumed to be isolated, i.e. no external tide was considered.

The process under the consideration, i.e. acceleration of stars to
velocities $>60\,\kms$ is quite rare. Therefore, we have integrated
several hundreds of different realizations ($N_{\mathrm{{\rm run}}}$)
of each model in order to obtain statistically relevant results.

\section{Results}

\subsection{Velocity distribution}

\begin{figure}[t!]
 \includegraphics[scale=0.45]{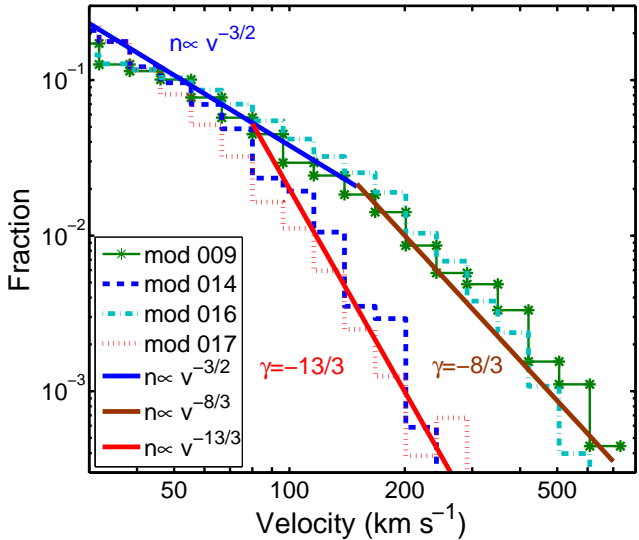}

\caption{\label{fig:vd}The velocity distribution of runaway stars ejected
from the simulated clusters with velocities larger than $20\,\kms$.
The runaways velocity distribution for cluster of different binary
separation distribution are shown; Dash-dotted and solid stair lines
are for a log-constant SMA distribution (models 009 and 016, respectively)
and dash and dotted lines are for uniform SMA distribution (models
014 and 017, respectively). See Table 1 for initial conditions of
all simulated clusters. The right segment of the straight lines (solid
brown and red) show the toy-model predictions for the runaway
velocity distributions, normalized to fit simulated data. Left segment
of the straight lines (solid blue) show power law distributions with
$n\propto v^{-1.5}$. Note that although the predicted slopes provide
the correct trends for the models, the best fitting (not shown) power
law slopes for the same high velocity regimes, $v>150\,\kms$, are
$-2.5$ and $-2.9$ for models 009 and 016, respectively (compared
with the predicted $\gamma=-2.7$); and for the other models, with
velocity regime $v>80\,\kms$, we get $-3.4$ and $-3.3$, for models
014 and 017, respectively (compared with the predicted $\gamma=-4.3$)}

\end{figure}

Figure~\ref{fig:vd} shows the velocity distribution of escaping
stars. We define escapers as those found at least 5 pc away from the host cluster at the end of the simulations, and have been ejected at the first 2.7 (35) Myrs of evolution for models 009 and 014 (016 and 017). The figure shows the velocity distribution compared
with the predicted distribution at the high velocity regime (when
taking the overall normalization to be a free parameter). Although
the analytic derivation does not account for the mass function, we
find that the velocity distribution of ejected stars in models 009 and
014 are also consistent with simple analytic formulation, and the slope
is determined mainly by the distribution of the semi-major axis of
the binaries (see fig. \ref{fig:vd}). The velocity distribution at
the lower velocity regime $v_{{\rm esc}}\approx20-150\,\kms$ is qualitatively
different, and appears to be independent of the distribution of the
binary separation. Comparison with a model where no binaries exist
show that all the bone-fide runaway stars arise from the existence
of binaries in the clusters. Although not the focus of this paper
it is interesting to note in passing that stars with velocities in
the intermediate velocity range $20-150\,\kms$ can not be explained
by single-single encounters alone; however, they also do not follow
the simple analytic approach described before, but rather appear to
follow a shallower distribution which is independent of the binary
separation distribution. This may be consistent with the recent results
of \citet{fuj+11} who studied runways from massive clusters in this
range of velocities. These may arise from the dynamical interactions
with a small number or even single very massive binaries in the cluster
(bullies), which dominate the ejection rate and are later on ejected
themselves from the clusters. The properties of runaways from such
interactions are independent of the overall binary population characteristics.
Our findings in Fig. \ref{fig:MF} showing ejection of very massive
runway binaries is consistent with this picture. Note, however, that
models 016 and 017 did not include massive binaries, which can play
the role of 'bullies'. Unfortunately, we did not keep data on specific
interactions leading to runaway ejections; we therefore can not directly
confirm or refute this scenario for the low velocity regime. We defer
such simulations and analysis to future work.

\subsection{Binarity of runaway stars}

\begin{figure}[t!]
\includegraphics[scale=0.45]{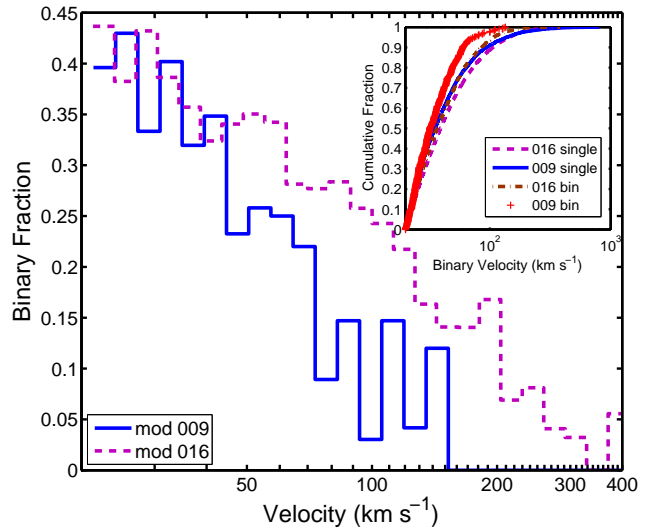}\caption{\label{fig:Vel-binaries}The binary properties of runaway stars. The
binary fraction of runway stars, as a function of the ejection velocity
in models 009 and 016. Inset: The cumulative velocity distribution
of runaway binary stars.}

\end{figure}

The binarity of runway stars is an important signature of the dynamical
ejection scenario. Previous studies of dynamically ejected runaway
stars suggested their binary fraction to be low relative to the binary
fraction of their parent stars in the cluster \citep{leo+90}. We
find a similar trend; Fig. \ref{fig:Vel-binaries} shows the velocity
distribution of single and binary runaways in models 009 and 014 of our
simulations (for which a realistic binary period distribution was
used), and the binary fraction of runaways as a function of their
velocity. The simulations results are consistent with theoretical
arguments; in a binary single encounter the binary receives a smaller
fraction of the kinetic energy, $M_{\star}/(M_{\mathrm{B}}+M_{\star})$
\citep{por+10}, it is therefore typically ejected at considerably
lower velocities. Note that encounters where the single star is much
more massive than the binary components likely lead to an exchange
of the massive components with one of the lower mass binary components,
therefore even such encounters would eventually lead to the binary
being the most massive component in the encounter, further contributing
to its low ejection velocity, compared to the ejection of single stars.
Binary-binary encounters may eject a binary to higher velocities,
however, these complex encounters may easily disrupt one of the binaries
and/or form higher multiplicity systems. A hard binary can be ejected
similar to the ejection of single star by a softer binary, however
its typical ejection velocity would be comparable to the orbital velocity
of the wide binary, again producing a bias toward low velocity ejections.
The overall binary fraction of runaways decreases significantly with
higher ejection velocities; and effectively no binary was ejected
at velocities higher than $\sim150\,(300)\,\kms$ in the 009 (016)
models (compare with the ejection velocities of single stars). Hyper-runaway
binaries are therefore not likely to be produced through the DES (see
also \citealp{per09a,bro+10} for discussions on this issue).

\subsection{The mass function and mass-velocity relation for runaway stars}

\begin{figure}[t!]
\includegraphics[scale=0.45]{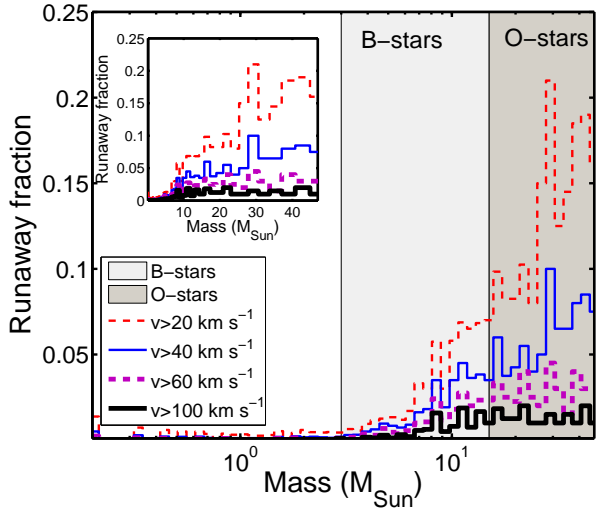}\caption{\label{fig:MF_frac} The fraction of runaways vs. mass with given
(lower limit) velocities for model 009. All stars below $4\, M_{\odot}$
are initially single stars, and their runway fraction might be non-realistic.
For OB stars $m>4\, M_{\odot}$ the runway fraction increases with
mass. This trend is consistent with the observed runaway fractions
of O stars to be $3-6$ times larger than the runaway B-stars fraction
(see \citet{sto91} and refs. therein). Note that even the most massive
stars are ejected as runaways; some of which are merger products,
with masses extending beyond the initial mass distribution (not shown). }

\end{figure}

In Fig. \ref{fig:MF_frac} we show the runaway fraction of stars.
As can be seen, OB runaways are relatively more abundant than lower
mass stars, with the O-star runaways fraction 2-3 times larger than
that of the B-stars, consistent with observations (see \citealt{sto91}
and refs. therein). 

The total fraction of runaways we find are comparable to, but systematically
lower ($\sim1/2$) than those reported by \citet{sto91}; this may
result from the limited time of the simulation; the lifetime of typical
O-stars could be 2-3 times longer than the simulation time (which
is comparable to the lifetime of the most massive stars in our simulations).
Obviously, many other simplified assumptions we use affect our theoretical
results, and may contribute to the difference. In particular the binary supernova scenario, which is not studied here would also contribute to the runaways population. 

\begin{figure}[t!]
\includegraphics[scale=0.45]{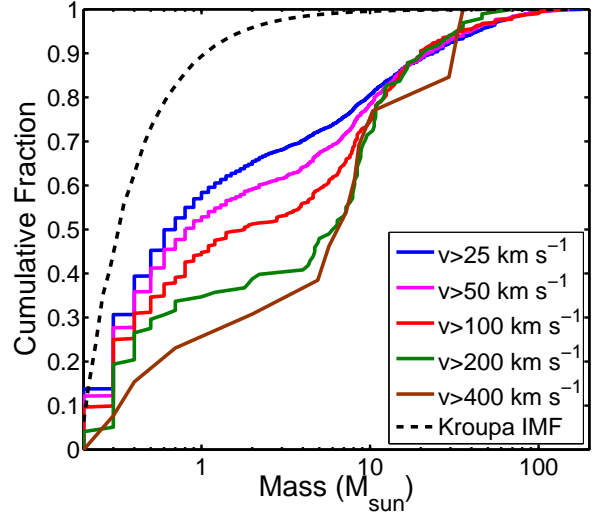}\caption{\label{fig:MF}The mass function of runaway stars in mod 009 as a
function of lower limit velocity. The distribution tends to be bi-modal,
with runways in the range $1-4$ M$_{\odot}$typically being least
represented in the population (compare with the shown IMF of all stars
in the initial simulated clusters). }

\end{figure}

In Fig. \ref{fig:MF} we present the mass-velocity distribution of
the ejected stars. We see again the trend of more massive stars having
higher runaway fraction (as reflected by the large fraction of massive
runaways compared with their fraction in the IMF); we note that even
the most massive stars are ejected as runaways. We also find that
the velocity distribution of more massive runaways tends towards higher
velocities than the lower mass stars, i.e. the fraction of more massive
stars increases with ejection velocity. In fact, we find that $\sim20$
$(10)$ \% of the runaway O-stars with $v>20$ $(100)\,\kms$ are
more massive than $40$ $M{}_{\odot}$, and $\sim5$ (3) \% of the
O-stars are with $v>20$ $(100)\,\kms$ are more massive than $80\, M_{\odot}$
(consistent with finding of very massive runaway stars; \citealp[e.g.][]{hoo+01}).
The latter are runaway merger products of two or more less massive
stars, since the adopted IMF of the clusters upper mass cut-off was
80 ${\rm M}{}_{\odot}$. Similar trends are found by \citep{ban+12} is their simulations of more massive clusters.

\subsection{Hyper-runaways and hyper-velocity stars}

As can be seen for both the two-mass models and the continuous mass
ones, the high velocity tail of the distribution follows a steep power
law. The ejection or the fastest runaways requires the close encounters
with binaries, the rate of which are dominated by the most dense clusters
(which are also the most massive ones, typically). Therefore, in order
to provide a basic estimate of the ejection of the fastest runaways
we consider here only the most massive clusters in the Galaxy. The
average number of hyper-runaways ejected with $v_{{\rm esc}}>300\,(>450)\,\kms$,
in our most realistic cluster, model 009, is $\sim0.2\,$ ($0.02$)
stars per cluster. Currently, $N_{{\rm c}}\sim10-20$ young ($<t_\mathrm{c}=10$
Myrs) massive clusters ($M>10^{4}\msun$) with cores comparable
to our simulated clusters exist in the Galaxy \citep{mur+10}. We
therefore expect the ejection rate of hyper-runways to be $\sim0.2\times N_{{\rm c}}\times t_{{\rm c}}^{-1}=0.2-0.4$
Myr$^{-1}$. For B-stars (of masses $\sim3-10$ M$_{\odot}$ such
as observed among the hyper-velocity stars in the Galactic halo),
the rate is $\sim2.5$ times lower. Over $100$ Myrs (comparable to
the propagation time for the observed HVSs in the Galactic halo),
we therefore expect to have of order 10-20 hyper-runaways in the galaxy,
but only one or two such stars with $v_{{\rm esc}}>400$. 
Currently $\sim20$ HVSs ($v_{{\rm esc}}>400$; though this is generally a good definition, note that a more accurate definition also depends on distance from the Galaxy, see \citealp{per+09}) have been observed in the Galactic halo, from which a total
of $\sim100$ B-stars (of $3-4$ $M{}_{\odot})$ are inferred to exist
in the Galaxy \citep[and references therein]{bro+06a,bro+09,bro11}, and
a few hundred lower velocity HVSs (so called bound HVSs; $275<v_{{\rm esc}}<400$ \citealp{bro+07}),
may exist in the Galaxy \citep{per+09}. Comparing the predicted and
observed numbers of HVSs, hyper-runaways from this scenario may contribute at most $1-2\,\%$  of the HVSs population and are unlikely
to have produced any of the currently {\it observed} HVSs in the Galactic halo (or at most one); though they may contribute a small fraction of the observed bound HVSs.

\section{Summary}

In this paper we explored the dynamical ejection of runaway stars
from the cores of massive stellar clusters. We presented a simple
analytical toy model to explain the velocity distribution of the fastest
runaway stars. We then used extensive N-body simulations of simplified
clusters composed of only two type of stars with masses 1 and 10 $\msun$
to characterize the velocity distribution of runway binaries; later
we also simulated clusters with realistic continuous Salpeter IMF.
We found the trends in velocity distribution to depend on the properties
of binaries in the cluster and be generally consistent with the analytical
toy models of multi-interactions with multiple binaries at the high
velocity regime $v\gtrsim80\,\kms$ (or $v\gtrsim150\,\kms$; depending
on the model). At the lower velocity regime, ejection velocities are
less dependent or possibly even independent of the binary properties,
suggesting a different channel for runaway ejections dominate this
regime (e.g. \citealp{fuj+11}).

We characterized the velocity distribution of runways from the continuous
IMF clusters, and discussed their mass function and binarity. We find
that the runaway fraction of stars increases with mass, consistent
with observations; and that very massive stars formed through collisions
in the cluster could be ejected as runaways. The binarity of runaways
stars decrease with increasing velocity, and is generally lower than
the binary fraction of their birth cluster. In particular, we find
the maximal velocity of binary runaways is limited to $<200\,\kms$.
We also find the mass function of runaways to be velocity dependent.
Although runaways are much more frequent amongst the massive stellar
population $>4$ $M_{\odot}$, we find that a large population of
low mass runaways should also exist. The ejection rate of hyper-runaways,
with velocities $>300\,\kms$ appear to be too low to explain a significant
fraction of the observed HVSs in the Galactic halo, and could at most
explain a small fraction of the observed bound HVSs. When combined with models for the runaways propagation in the Galaxy, our models can be used to predict the spatial distribution of runaway and hyper-runaway stars in the Galaxy \citep{bro+09b}, and could be constrained by future surveys.  

\acknowledgements{HBP is a CfA and BIKURA (FIRST) prize fellow. L\v{S} acknowledges
support of the the Czech Science Foundation via grant GACR-202/09/0772
and from the Research Program MSM0021620860 of the Czech Ministry
of Education.}

\bibliographystyle{apj}

\end{document}